\documentstyle[sprocl]{article}
\def\sla{\raise.15ex\hbox{$/$}\kern-.57em}

\def\Journal#1#2#3#4{{#1} {\bf #2}, #3 (#4)}

\def\NPB{{\em Nucl. Phys.} B}
\def\PLB{{\em Phys. Lett.}  B}
\def\PRL{\em Phys. Rev. Lett.}
\def\PRD{{\em Phys. Rev.} D}

\def\CMP{\em Comm. Math. Phys.}

\def\be{\begin{equation}}
\def\ee{\end{equation}}
\def\bea{\begin{eqnarray}}
\def\eea{\end{eqnarray}}
\begin{document}

\title{
\vspace{-5.0cm}
\begin{flushright}{\normalsize RUHN-00-3}\\
\end{flushright}
\vspace*{0.5cm}
EXACT CHIRAL SYMMETRY WITH A NON-PERTURBATIVE
CUTOFF.
\footnote{Invited talk at the international conference 
``Quark Confinement and the Hadron Spectrum IV'',
July 3 - July 8, Vienna, Austria.}
}
\author{H. Neuberger}

\address{Department of Physics and Astronomy, Rutgers University, 
Piscataway,\\ NJ 08855, USA\\E-mail: neuberg@physics.rutgers.edu}


\maketitle\abstracts{ The main ideas behind the new ways
to preserve chiral symmetries for lattice fermions are
presented. The discussion is focused on vector-like
fermions, the case relevant for lattice QCD.
}

Chiral effective Lagrangian techniques provide a succinct
and reasonably accurate description of low energy QCD~\cite{hol}.
This description is possible because it only reflects
the approximate chiral symmetries of QCD. Taking
the number of light quarks as three there is an
approximate global $SU(3)_L\times SU(3)_R$ symmetry.
The symmetry is broken explicitly, but weakly, by
the quark masses. There is no global $U(1)_L \times
U(1)_R$ because it is explicitly, but
strongly, broken by instantons. Only one abelian
global symmetry remains, a vectorial $U(1)_V$. 

The left and right ($L,R$) components of the symmetry 
group are best understood by thinking about Weyl fermions
as the fundamental building blocks of matter, rather than
Dirac fermions. One Dirac fermion is made up of one
left and one right handed Weyl fermion. While a mass term
couples the two fields directly, the interaction with
gluons does not, because gluons carry spin one. In four
dimensions anything but a bilinear fermion interaction
is strongly suppressed by a much higher energy scale.
Ignoring weak interactions we conclude that in the absence
of mass terms, independent global rotations between
different flavors of the right and left Weyl quark fields
leave the Lagrangian invariant.

Many theories that differ in the ultraviolet,
but with the same structure of global symmetries
will be described by structurally 
identical effective Lagrangians 
in the infrared, only the numerical 
values of the parameters would differ.
In particular, this class ought to include, to leading order
at low energy, also lattice regularizations of QCD,
provided only they keep the symmetry structure intact.
The lattice case is somewhat special because
the lattice breaks Euclidean Lorentz
invariance: as a result, at orders higher 
than leading order in
the low energy expansion, there will be differences between the 
effective Lagrangian applicable to the lattice and
the one applicable to QCD~\cite{rebeff}. (Also, in practical
simulations, Euclidean
Lorentz invariance is broken by the toroidal
boundary conditions one typically uses in a lattice 
simulation. These breakings also do not affect the
leading order term.)

The effective Lagrangian framework involves, in a direct
manner, such fundamental QCD quantities
as the pion and kaon decay constants and the
various current quark masses. This is very well
known in the continuum. The same story holds also
on the lattice, and, in principle, the above quantities
would play a central role in the lattice effective Lagrangian. 
The point I wish to stress is that this should happen
even relatively far from the continuum limit: Using
the leading order effective lattice Lagrangian on the lattice,
before the continuum limit is taken, there are natural
definitions of decay constants and quark masses. This
is not how these quantities are extracted by conventional
lattice methods, because these methods break explicitly
chiral symmetry and restore it only in the continuum limit.
There are then no natural effective Lagrangian definitions
of these quantities before the continuum limit is taken.

Based on the experience~\cite{higgs} with the lattice 
realization of the (potentially
strongly interacting) Higgs sector one expects
significant advantages for the numerical extraction of
these parameters (in particular the pion decay
constant) when an effective
Lagrangian description is applicable directly 
on the lattice. 

So, one would like the global symmetry features of QCD to
be incorporated into any non-perturbative approach,
but from 1973-1992 it was thought that lattice 
regularization was not intelligent enough to do that 
and there was no escape from having 
to first go close to the continuum limit before chiral 
symmetry became approximately correct. One could not
assure exact masslessness of quarks on the lattice 
so one could
never see the spontaneous symmetry breakdown effects
(like, for example, masslessness of pions for zero mass
up and down quarks) in isolation from the explicit
breaking effects induced by the small, but nonzero 
true quark masses. The entire concept of small quark masses
in which one can perturb would not apply to the lattice
without also requiring closeness to the continuum limit.

The main reason for the failure of the lattice  to preserve
chiral symmetries is that the lattice does too perfect a job
of preserving the continuum principle of minimal substitution
($p\rightarrow p_\mu +eA_\mu $ and its non-abelian generalization).
This property of the lattice regularization implies that
any exact global symmetry can be ``gauged'' turning it
into an exact local gauge invariance. Local gauge invariance
means that the theory depends on fewer degrees of freedom
than must be used to express the action in a form that employs a local
Lagrangian. But, the well known inevitability of anomalies~\cite{gaume} tells us that
when one tries to gauge certain kinds of chiral symmetries
the extra degrees of freedom do not totally and exactly decouple;
a certain class of actions (the Wess-Zumino actions) 
cannot be eliminated by local redefinitions, and keep these degrees
of freedom coupled to the rest. To avoid a paradox it must be
then that it is
impossible to explicitly preserve certain chiral global symmetries 
(including those we need for QCD) even only globally.
Indeed
when one tries to include such ``forbidden'' global chiral
symmetries one encounters all kinds of
difficulties. While the difficulties may take different forms,
the fundamental reason for their appearance is the existence
of irremovable Wess-Zumino actions in the continuum. In this
sense, the difficulties themselves can be viewed as universal. 
The decade between 1985 and 1995 saw a substantial,
albeit rather misguided, activity in trying to devise ways
around these obstacles. These efforts amounted
to an industry of failures which made many workers in the field
give up on the problem and become very skeptical about the
prospects of achieving exact chiral symmetry on the lattice. 
Thus, although the essence of a solution was available already
in 1993, it took six years more for a substantial fraction of the lattice
community to finally accept it. 

The main catalyst for the new development were two totally
independent papers, published during the same year, one by Frolov
and Slavnov~\cite{fro} and the 
other by Kaplan~\cite{kap}. 
Narayanan and I~\cite{ovl} synthesized
the main ideas of these papers into what became later to be known
as the ``overlap''. The basic idea can be loosely phrased
as hiding a Weyl fermion among an infinite number of Dirac
fermions. If we think about the Dirac fermions as pairs
of Weyl fermions we have an infinite number of left Weyl fermions
and an infinite number of right Weyl fermions. If the two infinities
are equal to each other the theory is vector like, but if they
differ by one we have an extra Weyl fermion. If indeed there
is this mismatch, any fermion number conserving
mass matrix we would introduce would be able to lift the masses
only of Dirac Weyl-pairs, leaving one unpaired zero mass 
Weyl fermion. 

To make this more concrete, consider the fermionic part
of a multi-flavor generalization of QCD,
\begin{equation}
\begin{array}{rcl}
{\cal L}_{\rm fermion} =\bar\psi {\sla {D}}(A) \psi +
\bar\psi {{1+\gamma_5}\over 2}
M\psi +\bar\psi {{1-\gamma_5}\over 2}
M^\dagger \psi \\
=\bar\psi {\sla {D}}(A) \psi +
\bar\psi_R 
M\psi_L +\bar\psi_L M^\dagger \psi_R
\end{array}
\end{equation}
and assume the mass matrix $M$ is infinite. How many
fermions are is determined by the operator $M$. (The proper
generalization of the concept of a finite matrix to
the infinite situation is to introduce some
Hilbert space in which $M$ acts as an operator.)
While saying how many flavors we have is meaningless,
the difference between the number of right handed
and left handed Weyl fermions is dynamically fixed
by the structure of $M$, so long the kernels of $M$
and $M^\dagger$ are finite dimensional, a very common
case in physics applications. The difference between
the dimensions of the two kernels is invariant under
suitably limited variations of the mass operator; it
is an ``index''. The index cannot change in perturbation
theory, so, if we arranged to have a surplus of one right 
handed Weyl fermion by picking a certain structure for $M$,
this massless degree of freedom will not disappear under
radiative corrections coming from the interaction with
gluons. Since the Weyl degree of freedom is unpaired
one cannot generate a mass term for it. Since the total
number of fermions is infinite, there is some uncertainty
about how accurately the Lagrangian defines the system.
Thus, one can expect to have exact global chiral symmetries,
and postpone the expected obstacles to the point when
one must give a precise meaning to the infinite number 
of flavors. 

The simplest explicit realization of a mass matrix
with unit index is by domain wall fermions. The Hilbert
space is taken as the space of square integrable
functions on the real line. It is possible to replace
the infinite real line by an infinite one dimensional
lattice, but this is a technicality, and probably not
a very useful one. A simple choice for $M$ is:
\begin{equation}
M=-\partial_s + \Lambda ~ {\rm sign} (s), ~~~~~~~~s\in(-\infty, +\infty).
\end{equation}
It has an index because
\begin{equation}
M\psi=0 \Rightarrow \psi=\psi_0 \propto e^{-\Lambda |s|},
\end{equation}
which is normalizable, but
\begin{equation}
M^\dagger \psi =0 \Rightarrow \psi \propto e^{\Lambda |s|},
\end{equation}
which is not.

Also, excepting $\psi_0$, all eigenstates of $M^\dagger M$ are
paired with those of $MM^\dagger$ and have energies of
order $\Lambda$. We now view $\Lambda$ as an additional
ultraviolet cutoff in our theory, and the entire infinite
set of massive Dirac fermions have masses of the order of this
additional ultraviolet cutoff. There is a finite and large
spectral gap at zero. This arrangement is stable under perturbations
of the mass matrix $M$ by finite norm operators because
the perturbation must cause shifts of order $\Lambda$
before it can move the zero energy state away from zero.
The main point is that it is the operator $M$ that gets perturbed,
so the perturbation of the operator $M^\dagger M$ is of a special type.
 
The system can be viewed as living in five (three) dimensions
when only four (two) are physical and the extra dimension
is seen only by the fermions. 
\begin{equation}
\bar\psi_R M\psi_L + \bar\psi_L M^\dagger \psi_R =
\bar\psi [\gamma_5 \partial_s + \Lambda ~{\rm sign} (s) ]\psi
\end{equation}
The gauge vector potentials do not see the extra dimension:
there is no $A_5$ ($A_3$) and all $A_\mu$ are $s$-independent.
Thus, the gauge fields act as the zero mode of the gauge field
in KK dimensional reduction.

The main advantage of this choice is that one can easily
interpret the fermion integral if one views the new
line variable $s$ as an Euclidean time, sort of a fifth
dimension only the fermions are aware of.
For a fixed gauge background the fermionic path integral 
one needs to give a precise meaning to has the action 
\begin{equation}
S_\psi=\int_{-\infty}^{\infty} ds \left [ \int_x L_\psi \right ]
\end{equation}
\begin{equation}
{\cal L}_\psi=\cases{ \bar\psi\gamma_5 [\partial_s + H(-\Lambda, A) ]\psi &
if $s < 0$\cr
\bar\psi\gamma_5 [\partial_s + H(+\Lambda, A) ]\psi &
if $s > 0$}
\end{equation}
The obvious formal interpretation is
\begin{equation}
\int d\bar\psi d\psi e^{S_\psi} = \langle -\Lambda , A| \Lambda , A \rangle
e^{-\infty [ E_+ (A) + E_-(A) ]}.
\end{equation}
$|\pm \Lambda , A\rangle$ are Fock states made out
by filling all states in the Dirac seas associated with
the single particle Hamiltonians $H(\pm \Lambda, A)$. 
The last factor is infinite, but gauge invariant. It arose
from integrating out an infinite number of massive
Dirac fermions. Therefore it is natural to discard it, as it
should not have any effect on the continuum limit.  
Then one obtains
a simple formula for the induced effective action, the
``fermion determinant''. This is the overlap: 
\begin{equation}
\langle -\Lambda , A| \Lambda , A \rangle ,
\end{equation}
where the states $| \pm \Lambda , A \rangle$ are 
Fock ground states of two
systems of noninteracting fermions, with single particle 
hermitian Hamiltonians
$H(\pm\Lambda, A)=\gamma_5 [ \sla{D}(A) \pm \Lambda ]$.

The two ground states are for two completely regulated
systems of non-interacting fermions moving in an 
arbitrary gauge field background. Thus, to calculate
the overlap one simply needs to diagonalize two
finite and explicitly known matrices.  These matrices
depend on the gauge background and transform by conjugation
when the background changes by a gauge transformation.
Therefore the eigenvalues are gauge invariant and the
eigenspaces transform covariantly. 
The gauge fields can be viewed as external parameters 
the two matrices depend on. The gauge field background 
consists of a collection
of all link matrices on a finite lattice, with toroidal boundary
conditions. 

Although the matrices are well defined, the bra and
ket representing the ground states have a phase arbitrariness.
Recall that we are still focusing on one multiplet
of right handed (say) Weyl fermions. Thus, an ambiguity is
necessary to permit anomalies to enter. Without the ambiguity,
complete finiteness and gauge covariance of expressions leading
to a chiral fermion determinant would have been impossible. 
The proper mathematical description of the overlap is not
that of a function over the space of gauge fields, but
as a line bundle over the space of gauge orbits. This
requires some more discussion, and some aspects relevant
to QCD will be covered later on. In the chiral case one
needs to pick some section through the line bundle built
out of combining all individual multiplets. The mathematics
is somewhat involved, but the bottom line is that one ends up~\cite{twod} 
being able to reproduce on the lattice such subtle physical phenomena
as, for example, composite massless fermions needed in order to comply
with 't Hooft's consistency conditions.

In the vector-like case, relevant to QCD, we have two bundles,
one associated to the left handed Weyl fermions and another
to the matched right handed Weyl fermions. Although the individual
sections are hard to choose, it is easy to see that any one choice
for one handedness has a matching choice for the other. Thus,
the combined bundle is trivial, and there is no ambiguity. The 
combined contribution of the left and right handed Weyl fermions
is given by a function over the space of gauge orbits. The
Dirac fermion determinant is given by a positive quantity:
\begin{equation}
\langle -\Lambda ,A | +\Lambda , A \rangle
\langle +\Lambda ,A | -\Lambda , A \rangle
=|\langle -\Lambda ,A | +\Lambda , A \rangle|^2 .
\end{equation}
So long we do not try to factorize the expression on the RHS 
back into two the two complex conjugate factors on the LHS, 
one for each handedness, there is no
ambiguity. 
But, the mere possibility to factorize the determinant
means that global chiral symmetries are still present
in some sense although  
these symmetries are not realized 
in a totally obvious way.
One way to make it explicit 
how the global chiral symmetries got hidden is
to first simplify
the expression for the vector-like chiral determinant on the RHS. 

This
is relatively easy, once one observes that the 
two Fock ground
state rays are completely defined (in the absence 
of accidental 
degeneracies - see below) if we know only the 
linear subspaces
of the individual single particle systems that 
correspond to
negative energies. In other words, we do not
need to choose individually exact eigenstates
of $H(\pm \Lambda , A)$ corresponding to each
negative eigenvalue: we only need to define 
the subspace spanned by all eigenstates corresponding
to negative eigenvalues. Any orthonormal basis
in this subspace can be used to construct the 
Fock ground state, not just the one made out of
single particle eigenstates. 
Thus, there must exist a way 
of expressing
the Dirac fermion determinant in terms of the two 
projectors
onto these subspaces. 

In the concrete case of 
the single particle fermion
lattice dynamics being 
governed by the Wilson Dirac operator, 
one easily
can prove that nothing is 
lost by making one of the subspaces,
and the projector on it, trivial and 
gauge field independent.
For the other subspace the 
dependence on gauge field is crucial
and nontrivial. Linearly related to the projector is the
sign function of the appropriate single particle Hamiltonian
matrix. The sign functions of $H(\pm \Lambda, A)$
play a central role in the overlap Dirac operator
to be introduced below. When the simplifying case
of replacing $H(+\Lambda, A)$ by $H(+\infty, A)$ 
is chosen, as mentioned above, ${\rm sign} (H(\Lambda, A))$
becomes trivial, being given by $\gamma_5$. 
The basic identity~\cite{do} 
that applies in this case is
\begin{equation}
|\langle \infty | -\Lambda, A \rangle |^2=
\det D_o,
\end{equation}
where,
\begin{equation}
D_o = {{1+\gamma_5 {\rm sign}~H_W (m, U)}\over 2}.
\end{equation}
Here, I reverted to more standard notation for $H$,
using $m$ instead of $-\Lambda$, and the link variable
symbol $U$, instead of $A$. The subscript indicates that
we use Wilson's form. The parameter $m$ is somewhere 
between $0$ and $-2$, but neither $0$ nor $-2$. 
Theoretically, the simplest case to analyze is $m=-1$.
%%It easy to check that the formula is correct:

Had we not introduced the one 
(merely technical) simplification
for one of the subspaces we would 
have made no commitment to
the kind of regularization we are using. 
To be concrete however,
we need to go sooner or later to the lattice. Here we face
the well known problem of nontrivial topological sectors.
 
For a while, after the spectacular emergence
of ``instanton'' physics, 
it was believed that there was no clean way in 
which the space of lattice gauge field configurations
can reproduce the property of 
the same space in the continuum, 
namely that it is disconnected into 
sectors uniquely identified
by one signed integer. Indeed, the 
space of lattice configuration
is just a finite product of group 
factors and hence glaringly 
connected. But, as first shown by Phillips 
and Stone~\cite{phil}, following
some work by L{\" u}scher, one 
can slice up the lattice based
space by removing subsets of zero measure 
(relative to the natural measure of a product of
one factor of Haar measures per link, times
any positive smooth functional of the gauge fields)
and this division approximates the continuum
situation in a sense that can be made precise. 
On any finite lattice, the range
of topological numbers that can be 
accurately represented is limited
to a finite segment, as expected. 
While the details of the slicing-up
are not universal, all such 
divisions will agree in terms of
the topological number of a gauge 
background that is, in some
precise sense, smooth enough. The smoothness criterion is
gauge invariant: any two backgrounds 
related by a lattice gauge
transformation are regarded as 
having the same amount of smoothness.

When massless fermions are present, 
the behavior in each topological
sector is characterized by a different multilinear object
of fermion fields that acquires a 
non-zero expectation value upon 
fermion integration. This object 
is gauge invariant, and
$SU(3)_L\times SU(3)_R\times U(1)_V$ invariant. At zero
topological charge the object is unity, but at nonzero 
topological charge it is non-trivial and its non-zero
expectation value (after averaging over gauge backgrounds) 
breaks $U(1)_A$. Since 
the behavior in the different sectors
is so distinct, when lattice gauge fields are smoothly
deformed from one sector 
into another, something singular must happen
somewhere along the way. What happens is that the sign
function becomes ill defined. It is ill defined
for gauge field backgrounds 
where the many-body ground state 
is degenerate. This happens precisely  
where the single particle
Hamiltonian matrix has an exact zero eigenvalue. Clearly,
this happens only on a subset of gauge fields of zero measure.
Moreover, since the eigenvalues are gauge invariant, the
criterion is gauge invariant. The sign function is 
defined only for gauge backgrounds 
for which there are no zero
modes to the single particle Hamiltonian. Thus, the space
of gauge orbits is partitioned, just like in the 
Phillips and Stone scenario. 

The division of the space of gauge 
orbits is done by using fermions.
In this sense, the overlap provides, as a side result, a
``fermionic Phillips and Stone construction'' of topology
on the lattice. The fermionic character of this construction
makes it trivial to go beyond Phillips and Stone, towards a
lattice version of the 
Atiyah-Singer theorem itself. It becomes
almost tautological: Since we 
defined lattice gauge field topology
by the fermions the relation to fermions is built in. 
To be sure, complete contact with continuum is made
only after one makes it precise what the requirements
of the lattice one needs to use for a given
continuum configuration are, in order to ensure that
the fermionic topological number given by
the overlap coincides with the topological number
of the continuum gauge field configuration. 
If we just move one important step beyond the Atiyah-Singer
theorem to include the essential physics, nothing is 
even remotely tautological
any more. Just like 't Hooft showed 
in the continuum, the nontrivial
topology of the gauge background implies explicit $U(1)_A$
breaking by expectation values 
of `` 't Hooft vertices'', on the lattice
too we see that in nontrivial 
sectors we have non-zero 't Hooft
vertices. This is the main physics 
application of the Atiyah-Singer theorem,
and this is what really matters in QCD. 
It is this feature that
one needs to reproduce on the lattice if 
the latter is to solve
the ``$U(1)_A$ problem'' in the way discovered by 't Hooft.
To summarize, the overlap produces a good definition
of topological charge, via the formula:
\begin{equation}
Q_{\rm top} (U)=-{1\over 2}
Tr \left [ {\rm sign } (H_W (m, U) )
\right ]
\end{equation}

The trace operation involves a sum over sites and over
spinorial and color indices. The sum over the sites
is over a local quantity, a lattice version of the
continuum topological density. This can be shown
to hold using perturbation theory or more sophisticated
means, so long the lattice is fine enough in some precise
sense. When $Q_{\rm top} (U)$ is non-zero the number
of negative energy states of $H_W (m, U)$ differs
from one half the total number of states. As a result
the fermionic determinant is trivially zero, since by
our identity $\det D_o$ is given by the overlap of two
orthogonal states. Moreover, we know how many creation operators
need to be inserted between the two states making
up the overlap to make it non-vanishing. These operator
insertions naturally correspond to individual Grassmann
factors making up 't Hooft's vertex. From the
point of view of the conserved fermion number 
operator in the auxiliary Quantum Mechanical problem
producing the overlap we need an insertion that
carries a specific fermion number in one factor of
the overlap and minus that fermion number in the
other factor. Thus $U(1)_V$ is conserved but $U(1)_A$
is not, because it measures the difference between the
fermion numbers in the two factors.

Having discussed how the one 
unwanted chiral symmetry is avoided
by the lattice, it remains to be seen, 
explicitly, how the remaining
global chiral symmetries are preserved by the lattice. 
In other words, we wish to see where chiral invariance
is hidden in the action
\begin{equation}
{\cal L}_{\psi,~V} = \bar\psi D_o \psi
\equiv \bar \psi {{1+V}\over 2}\psi,
\end{equation}
where $V=\gamma_5\epsilon \equiv \gamma_5 {\rm sign} (H_W(m,U))$
is unitary and obviously obeys
\begin{equation}
\gamma_5 V \gamma_5 = V^\dagger = V^{-1} .
\end{equation} 
$D_o$ was obtained integrating out heavy degrees of freedom
from a somewhat formally defined, but explicitly chiral
(chiral because there was no direct coupling between the
left and right Weyl components of the Dirac fermion),
system. This is very similar to the approach of Ginsparg and Wilson~\cite{gw} 
with the single conceptual difference that the heavy fermions integrated
out by Ginsparg and Wilson corresponded to the high momentum modes
of a continuous Dirac field. But, in both cases, on the way to a concrete
action one hides chirality. The remnant is reflected by the 
``$\gamma_5$-hermiticity'' of the unitary matrix $V$ which defines
$D_o$. This property of $V$ is essentially equivalent to the Ginsparg-Wilson relation. 

The fermionic propagator on internal fermion lines must generate
the determinant, so has to be
\begin{equation}
G={2\over{1+V}}
\end{equation}
A simple calculation now shows that
\begin{equation}
\gamma_5 G \gamma_5 = 2 - G,
\end{equation}
implying that 
\begin{equation}
G_{\chi}\equiv G-1 = {{1-V}\over{1+V}} = - \gamma_5 G_{\chi}\gamma_5
\end{equation}
and hence $G_\chi$ anticommutes with $\gamma_5$. 
In a pure gauge background
the propagator $G_{\chi}$ vanishes at the locations of the doublers
($V=1$) and has the expected pole at zero momentum ($V=-1$). 
Thus, the free action $G_{\chi}^{-1}$ (with $U\equiv 1$) has poles
at the doublers, just as suggested years ago by Rebbi~\cite{rebpol}. It is a nonlocal
operator, but it is chiral. One cannot use $\det [G_{\chi}^{-1}]$ (for
arbitrary $U$) as
the fermion determinant, as shown by Pelissetto: the poles at the doublers
represent non-local couplings that have measurable effects in the
continuum limit~\cite{pel}. So, $G_{\chi}$ is inadequate as a propagator on
internal fermion lines and one must use
$G$ which has only the physical pole
and no zeros. However, $G_\chi$ is perfectly adequate on external
fermion lines, where it would provide the needed chiral identities
relating various fermionic correlators in a fixed gauge background. But,
is it consistent to have different propagators on internal and 
external fermions lines ? The answer is positive, as is easily
seen by introducing an auxiliary fermionic variable $\xi$. $\xi$
has a gauge invariant action ($\xi$ transforms the same way as
$\psi$) designed to make no contribution to the fermion
determinant but subtract the identity from the propagator of $\psi$,
$G$: 
\begin{equation}
{\cal L}_\xi = -\bar\xi\xi .
\end{equation}
We now declare that the ``physical'' fermionic fields $
\bar\psi_{\rm ph},\psi_{\rm ph}$ are given by:
\begin{equation}
\psi_{\rm ph} = \psi + \xi, ~~\bar\psi_{\rm ph} =\bar\psi+\bar\xi .
\end{equation}
Clearly, the propagator 
$\langle \psi_{\rm ph}\bar\psi_{\rm ph}\rangle$ is $G_\chi$, which is
chiral. So, we use only $\psi_{\rm ph}$ fields when we construct
quark operators whose matrix elements we wish to evaluate and these
operators obey all chiral identities we know from the continuum.

Time has come to get more technical. One may wonder whether it
would be possible to change the measure of integration over
the link variables in such a manner that the space of accessible
gauge field configurations would fall into  topological sectors in just 
the way needed for $H_W (m, U)$ to never have zero eigenvalues.
The answer is positive, because of the following rigorous inequality~\cite{bnd}:
\begin{equation}
\left [ \lambda_{\rm min} (H_W^2 (m,U))\right ]^{1\over 2} \ge
\left [ 1-(2+\sqrt{2}) \sum_{\mu > \nu } \epsilon_{\mu\nu} \right ]^{1\over 2}
-|1+m|
\end{equation}
where
\begin{equation}
||1-U_{\mu\nu}(x)|| \le \epsilon_{\mu\nu}.
\end{equation}
Here $U_{\mu\nu} (x)$ is the parallel transporter round one elementary
plaquette, with a corner at site $x$ and extending into the positive $\mu$
and $\nu$ directions.
We could write a local gauge action that puts zero probability on any
gauge field configuration that has any plaquette farther in norm from
unity than a given (small) amount and thus assure that no gauge configuration
that could produce a zero eigenvalue to $H_W(m,U)$ is allowed. This
way we are cutting out a very large piece of the total space of gauge fields,
much more than is really needed. But, the point of principle, albeit
of rather academic interest,  is that
the criterion is enforceable by an acceptable local gauge action. 
 
What really is of
practical relevance is how one should simulate the rather nontrivial
sign function of the hermitian Wilson Dirac operator~\cite{prac}. There is no space
here for a detailed review. In a nutshell the situation is as follows:
There are two somewhat distinct ways to create a good numerical approximation
of a system containing an exactly massless quark. In the first way,
which goes under the name of ``domain wall fermions'', one uses
a discrete ``fifth dimension'' $s$, and keeps it of finite extent.
This yields a system containing many 
heavy fermions, and one very light fermion.
It turns out that numerically one needs 
to go to large extents in $s$ in order 
to make the approximation work. The other way goes under the name
of ``overlap fermions'' and is based on direct truncations of integral
representations of the sign function. Both ways encounter numerical
difficulties when the spectrum of $H_W(m,U)$ gets too close to zero,
because the sign function has to jump by a finite amount when the sign
of an eigenvalue switches. While differences 
of the order of a factor of 2 or 3
cannot be ruled out, 
the bottom line conclusion, 
based on several recent efforts, is
that numerically the two methods 
are similar in cost. 

There is little doubt that overlap fermions are theoretically cleaner,
and therefore more adapted to analytical calculations, something
that is always needed when contact with continuum is sought after.
Therefore, barring some unforseen numerical development, I expect
domain wall fermions to be replaced by overlap fermions in QCD
simulations that use the new advances on realizing chiral symmetry
on the lattice. However, a large amount of computational resources,
by today's standards, has already 
been invested into domain wall 
fermions (with results of a somewhat
mixed quality), and judging by the 
history of the entire subfield of 
lattice chiral fermions, 
although we already know now that overlap fermions
would be better, it might 
take quite a few more years
for the switch to overlap fermions
to occur on a large scale. 

Let me end on a more positive note: Unlike in many cases in 
theoretical Physics, when hard 
problems often get redefined and shifted around,
the problem of lattice chirality has been truly solved. Eventually,
as a result of this, lattice field theory will change substantially and become more
effective for QCD. Moreover, lattice field
theory might be able to tackle chiral gauge
theories and provide some reliable
non-perturbative information on 
this extremely important 
class of field theories. 
I even believe that these developments will influence particle Physics
as a whole, because the difficulty to naturally produce a low energy
chiral gauge theory, without starting from one at higher energies,
transcends the lattice in its relevance.

\section*{Acknowledgments}
This research was 
supported in part by the DOE under grant \#
DE-FG05-96ER40559. I wish to thank the organizers of Confinement IV
for the invitation to participate, for support, and for the hospitality
extended.

\end{document}